\documentclass[aps,prl,amsmath,amsfonts,twocolumn]{revtex4-1}%
\usepackage{mathtools}
\usepackage{verbatim}
\usepackage{hyperref}
\usepackage{color}
\usepackage{graphicx}
\usepackage{braket}
\usepackage{amsthm}
\usepackage{amssymb}
\usepackage{ulem}
\usepackage{tikz}

\newcommand{\IK}{\color{red}}

\newcommand{\old}{\color{black}}

\begin{document}

\title{Emergent geometry and path integral optimization for a Lifshitz action}

\author{A. Ahmadain and I. Klich}
\affiliation{Department of Physics, University of Virginia, Charlottesville, Virginia 22903, USA
}
\begin{abstract}
 Extending the background metric optimization procedure for Euclidean path integrals of two-dimensional conformal field theories, introduced by Caputa et al. \cite{caputa2017anti,caputa2017liouville}, to a $z=2$ anisotropically scale-invariant $(2+1)$-dimensional Lifshitz field theory of a free massless scalar field, we find optimal geometries for static and dynamic correlation functions. For the static correlation functions, the optimal background metric is equivalent to an AdS metric on a Poincare patch, while for dynamical correlation functions, we find Lifshitz like metric. This results suggest that a MERA-like tensor network, perhaps without unitarity, would still be considered an optimal background spacetime configuration for the numerical description of this system, even though the classical action we start with is not a conformal field theory.
\end{abstract}


\maketitle
An important quest of many body physics is the search for efficient variational characterizations of correlated quantum systems. (for a review see, e.g., \cite{Orus2014}). A class of tensor network states, particularly geared towards the description of scale-invariant systems, are called the {\it multi-scale entanglement renormalization ansatz} (MERA) \cite{Vidal2007ER,Vidal2008MERA}. MERA is used to represent approximate ground states of 1D quantum spin chains at criticality described by 2D conformal field theory (CFT)\cite{pfeifer2009entanglement}. The scale-invariance of the MERA network turned out to also play a special role in connecting it to holographic duals in the sense of the AdS/CFT correspondence~\cite{swingle2012entanglement}. Here, the bulk of a MERA network can be understood as a discrete realization of 3D anti-de Sitter space ($ AdS_{3} $), identifying the extra holographic direction with the renormalization group (RG) flow in the MERA~\cite{swingle2012entanglement}.

The ground work for the connection between continuous tenor networks~\cite{haegeman2013entanglement,nozaki2012holographic,miyaji2015continuous} and path integral optimization for $AdS_{3}/CFT_2$ was initially laid out in~\cite{miyaji2017path, caputa2017anti,caputa2017liouville}. Recent work on the relationship between path integral optimization, different types of CFTs and complexity can be found in~\cite{sato2019does,jafari2019path,ghodrati2020complexity,caputa2021geometry}. Motivated by the procedure of tensor network renormalization in~\cite{Evenbly2015TNR}, %where the path integral is first discretized into a lattice and then mapped into a tensor network which turns out to be a MERA, it was
Caputa et. al  \cite{caputa2017liouville,caputa2017anti}, reinterpreted this connection as optimization of the background metric in the space of path integrals.
%took a step further in studying this relationship from the viewpoint of optimizing Euclidean path integrals that represent the ground state wave functional of two-dimensional CFT 
Starting with flat Euclidean metric with a UV cutoff, they argued that their optimization procedure amounts to minimizing the Jacobian of the scale transformation for the path integral measure. In the conformally flat gauge, this translates to solving the equation of motion of the Liouville effective action from which they find that the $AdS_3$ metric a Poincare patch $H_2$ naturally emerges. This new approach is very appealing, as it suggests a concrete procedure connecting the AdS/CFT correspondence with numerical approaches to many body systems, such as the MERA tensor network \cite{Vidal2007ER,Vidal2008MERA,swingle2012entanglement,Evenbly2011TNGeometry}.

\begin{figure}
	   \includegraphics[width=1.0\linewidth]{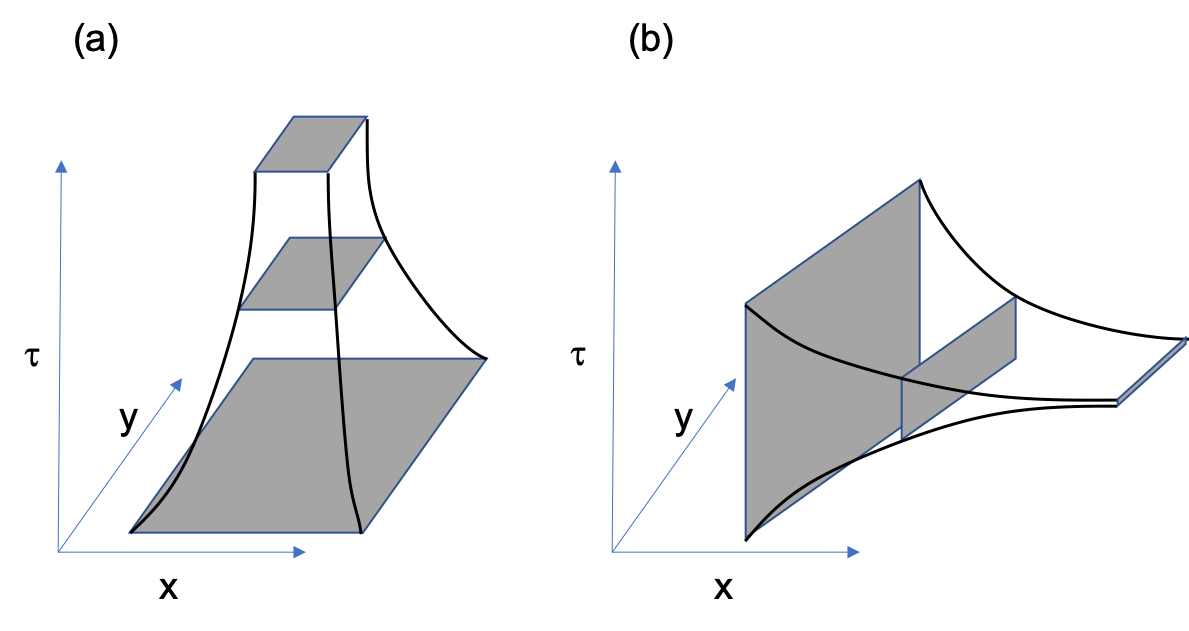}
	    \caption{The two geometries emerging for the quantum Lifhsitz model. (a) An $AdS_3$-like geometry arises when considering equal time correlation functions and (b) A Lifshitz metric that is optimal for computing correlation functions with a temporal component.} \label{fig:AdSLifshitz}
	\end{figure}

In this paper, we extend the idea in \cite{caputa2017anti,caputa2017liouville} to a \textit{non-relativistic} field theory, specifically to a $z=2$ anisotropically scale-invariant $(2+1)$-dimensional Lifshitz field theory of a free massless scalar field and show that the procedure can be successfully applied in systems of interest beyond a CFT. We show how natural geometries arise from the path integral optimization procedure. Our results are illustrated in Fig.  \ref{fig:AdSLifshitz}

%With proper rescaling of the Euclidean time coordinate, we show that this optimal background geometry takes the form of an $AdS_3$ metric on $H_2$ at the expense of starting with an action that breaks Euclidean time translation symmetry in the holographic direction.

The quantum Lifshitz model is a canonical example of a 
(2+1)-dimensional Lifshitz field theory known \cite{ardonne2004topological}. This model describes a free massless scalar field with dynamical scaling exponent $z=2$ and represents an important example of a conformal quantum critical point. Different aspects of this theory have been studied and analyzed in \cite{ardonne2004topological,fradkin2006entanglement,isakov2011dynamics}. For example, it emerges as the scaling limit of the square lattice quantum dimer model \cite{isakov2011dynamics}. 
%Non-relativistic Weyl anomalies in general have also been studied extensively analyzed in the litarature. [List all refreneces.] 

%We emphasize, however, that our main focus in this paper is \textit{not} to calculate the Weyl anomaly of this (2+1)-dimensional model.
%Here we use the Weyl anomaly to obtain an optimal metric for the quantum  Lifshitz field theory. 
%As in [Caputa], we carry out our heat kernel calculation in a fixed  gauge, the flat gauge and with the assumption that the conformal factor only depends on the temporal coordinate $t$. The direct result of this calculation is deriving an equation of motion for this conformal factor after which we proceed to solve. We will have more to say about these steps later in the paper.

%\red{Would like to put some TN background here or relegate to the end of the paper when you compare to your work with Raf and Evenbly?}

\old
The quantum Lifshitz Hamiltonian \cite{ardonne2004topological} of a $z=2$ theory of a massless scalar field $\phi(t,x)$ in $2+1$ dimensions is given by 
\begin{eqnarray} \label{Hamiltonian}
H=\int d^2 x ~\{{\pi_{\phi}}^2+({\Delta _s\phi})^2\} ~.
\end{eqnarray}
The Euclidean action of the field $\phi(t,x)$ coupled to a \textit{background} metric $g_{ij}$ is given by 
\begin{eqnarray} \label{ClassicalAction}
S = \int d^2x \text{dt} N\sqrt{h}\left(N^{-2}\left(\partial _t\phi \right){}^2+ 
\left(\Delta _s\phi \right){}^2\right) ~,
\end{eqnarray}
where $\Delta _s$ is the \textit{spatial} Laplace-Beltrami operator
\begin{eqnarray}
\Delta _s=\frac{1}{\sqrt{h}}\partial _ih^{\text{ij}}\sqrt{h}\partial _j ~,
\end{eqnarray}
and $h_{ij}$ is the spatial component of the background metric \footnote[2]{In terms of the ADM metric commonly used in the literature, $g_{tt} = \frac{1}{N^2}$}.
\begin{eqnarray} \label{BGMetric}
\text{ds}^2=N^2\text{dt}^2+h_{\text{ij}}\text{dx}^i\text{dx}^j ~.
\end{eqnarray} 
where $N$ is called the lapse function.
The action in (\ref{ClassicalAction}) is invariant under the following foliation-preserving diffeomorphism transformations 
\begin{equation}\label{FPD}
t\mapsto \tilde{t}(t), \quad  \pmb{x} \mapsto \tilde{\pmb{x}}(\pmb{x})
\end{equation}
%\begin{equation}\label{FPD}
%t\mapsto \tilde{t}(t), \quad  x^i %\mapsto \tilde{x}^i(\vec{x})
%\end{equation}
where $\pmb{x}=\{x^1, x^2, x^3\}$, and anistoropic Weyl scaling transformations 
\begin{eqnarray}
N\rightarrow  e^{z \sigma }N\text{   };\text{
}h_{\text{ij}}\rightarrow  e^{2\sigma }h_{\text{ij}}~, \text{    }i,j\neq t\text{  }~.
\end{eqnarray} 
As stated before, in \cite{caputa2017anti}, such a starting point led, via path integral optimization, to an AdS metric. 
The path integral optimization suggested in \cite{caputa2017anti} looks for the extremal measure over all choices of the gauge $\sigma$, due to the Weyl anomaly in the model. Here we use the same structure, though with the anisotropic Weyl scaling appropriate. 

Here we ask the following question: what is the optimal geometry associated with a path integral computation of correlation functions in the quantum Lifshitz model? In contrast to the CFT case, due to the non-relativistic nature of the model, equal time correlation functions and dynamical correlation functions should be treated differently. Indeed, we find two separate geometries associated with the optimal calculation, described in Fig. \ref{fig:AdSLifshitz}. For equal-time correlation functions, we consider Weyl transformation which are translationally invariant in space, but not in time, Fig. \ref{fig:AdSLifshitz}(a), covered by case (1) below.

Consider dynamical correlation functions on the other hand. To find the  optimized geometry to describe two point functions, such as, say, $\langle \phi(t,r)\phi(r',t)\rangle$, we can choose the spatial axis $r-r'$ to be in the $y$ direction, due to spatial rotational invariance of the model. We concentrate therefore on the computation of the description of the state in the $t,y$ plane, and thus choose a Weyl scaling which is homogeneous in $t,y$, but can depend on the third coordinate $x$, Fig. \ref{fig:AdSLifshitz}(b) as explained in case (2) below.

Of particular interest to us in this paper, is the Weyl anomaly of this model which has first been computed holographically in \cite{griffin2012conformal} and by Baggio et al in \cite{baggio2012anomalous} using heat kernel expansion and the holographic renormalization methods in \cite{baggio2012hamilton}. In \cite{Arav:2014goa} \cite{arav2017lifshitz}, Lifshitz Weyl anomalies have been computed cohomologically in different dimensions and for different values of the dynamical scaling exponent $z$. In \cite{barvinsky2017heat}, the heat kernel expansion has been generalized to calculate effective actions and Weyl anomalies for Lifshitz field theories. A general framework for computing one loop effective action for Lifshitz theory via heat kernel coefficients has been presented in several places, see e.g. \cite{nesterov2011gravitational,barvinsky2017heat}. 

We note that in contrust with \cite{caputa2017anti}, here, We do not start from the quantum effective action and then derive the equation of motion as they do but rather directly compute the variation in the Lifshitz effective action due to an infinitesimal transformation of the Weyl transformation parameter $\sigma$. 
Our starting point is a flat metric, deformed by a Weyl scaling, therefore $\sigma$ carries the entire information on the metric in the space of metrics we explore. We compute the variation of the effective action explicitly utilizing the particular structure of our metric and finally obtain differential equations for the scaling factor $\sigma$. Concretely, we compute the variation of the one loop effective action under $\sigma\rightarrow \sigma+\delta\sigma$. In this case,
\begin{eqnarray}\label{EffectiveActionVar}
    \delta W[\sigma]={1\over 2}\int d\pmb{r} \delta\sigma(\pmb{r}) \bra{\pmb{r}}e^{-\epsilon \rho D}\ket{\pmb{r}} ~,
\end{eqnarray}
where $\pmb{r}=(\pmb{x},t)$,  $\rho(\pmb{r})=\frac{1}{\sqrt{g(\pmb{r})}}$, $\epsilon$ is the infinitesimal heat kernel "time" parameter, and $D=-{1\over N\sqrt{h}}\partial_t N^{-1}\sqrt{h}\partial_{t}+{1\over N}\Delta_{s}N\Delta_{s}$ \cite{baggio2012anomalous}. In our system we fix our gauge so that $N=e^{2\sigma}$, $h_{ij}= N \delta_{ij}$. In this case we have:
\begin{eqnarray}
    D=\left(-\partial _t^2+\left(\partial _x^2+\partial
   _y^2\right){}^2\right).
\end{eqnarray}
%and $c$ is the anomaly coefficient associated with the action in (\ref{ClassicalAction}). 
We note that upon varying $\sigma$ we have $\delta D=-4\delta \sigma D$.
The $\epsilon\rightarrow 0$ behavior of (\ref{EffectiveActionVar}) is dominated by the short distance behavior of the heat kernel $\bra{\pmb{r}}e^{-\epsilon \rho D}\ket{\pmb{r}}$. 

Now, as promised, we specialize to cases where, $\sigma$ depends either on the time coordinate $t$ alone, or on one of the spacial coordinates, say $x$. 
Denoting $\rho = e^{-4 \sigma}$, we expand $\rho$ close to a given point ${\pmb{r}_0}$ ,
\begin{equation}
\rho\left(\delta \pmb{r}+\pmb{r}_0\right)=\rho_0 +\delta \rho,
\end{equation}
where $\rho_0=\rho(\pmb{r}_0)=\frac{1}{\sqrt{g(\pmb{r})}}|_{\pmb{r}=\pmb{r}_0}$.

To obtain the variation we carry out a second order perturbation calculation of the heat kernel, using :
\begin{eqnarray}\label{pert_series}&
e^{-\epsilon  \left(\rho _0+\delta \rho \right)
   D}=e^{-\tilde{\epsilon } D}\!-\!\frac{1}{\rho
   _0}\int _0^{\tilde{\epsilon
   }}e^{-\left(\tilde{\epsilon }-s\right) D} \delta \rho  D e^{-s D}\text{ds}\\ \nonumber &+ \frac{1}{\rho _0^2}\int
   _0^{\tilde{\epsilon }}\text{ds} \int
   _0^s\text{ds}_1e^{-\left(\tilde{\epsilon
   }-s\right) D} \delta \rho  D e^{-\left(s-s_1\right) D} \delta \rho
    D\text{  }e^{-s_1 D}
\end{eqnarray}
where $\tilde{\epsilon }=\rho _0\epsilon$. We assume that the operator $D$ is diagonal in momentum, and that $\delta\rho$ depends on a single coordinate such as $x$ or $t$ and has an expansion:
\begin{eqnarray}
\delta \rho =\Sigma _{m=1} c_m (x-x_0)^m
\end{eqnarray}
Explicitly evaluating the heat kernel through second order perturbation series in $\delta\rho$, we find that the leading (in $\epsilon$) contributions to $\delta W$ up to two derivatives are given as
\\
(1) $\sigma=\sigma(t)$. In this case:
\begin{eqnarray}\label{dW_equal_time}
\text{$\delta $W}=\frac{1}{2}\int
   \text{dtd}^2x \delta \sigma 
   \left(\frac{e^{4 \sigma }}{16 \pi
    \epsilon }-\frac{1}{24 \pi
   }\frac{d^2\sigma
   }{\text{dt}^2}\right)
\end{eqnarray}
(2) $\sigma=\sigma(x)$. In this case, the leading in $\epsilon$ contributions, read:
\begin{eqnarray}\label{dW_dynamical}
\text{$\delta
   $W}=\frac{1}{2}\int
   \text{dtd}^2x \delta \sigma
   \left(\frac{e^{4 \sigma
   }}{16 \pi \text{  }\epsilon
   }-\frac{e^{2 \sigma
   }(
   \left(\frac{\text{d$\sigma
   $}}{\text{dx}}\right)^2+\frac{d^2\sigma
   }{\text{dx}^2})}{12 \pi
   ^{3/2} \sqrt{\epsilon
   }}\right)
\end{eqnarray}

%\IK In \cite{nesterov2011gravitational}, the gravitational quantum effective action for a $d$-dimensional Lifshitz scalar field theory has been calculated using the heat kernel expansion in momentum space. It is important to note, however, that the curved spacetime Lifshitz operator used in \cite{nesterov2011gravitational} is slightly different than the one we use in this paper. We note that the terms above correspond to the heat kernel coefficients $b_0,b_2,b_4$ in the expansion:
%\begin{eqnarray}
%\langle \pmb{r}|e^{-\epsilon \delta \rho D}|\pmb{r}\rangle=\sum_{n=0}^{\infty} b_n(\rho D)\epsilon^{(n-d-z)/2z}
%\end{eqnarray} 
%specializing to $d=z=2$. Note that $b_1,b_3=0$ as anticiated e.g. in \cite{barvinsky2017heat} ({\bf actually they say $b_{n}=0$ for $n\in 2\mathbb{N}+1$, so I am not sure $b_1$ is covered}.\old

{\bf Optimized geometry for equal time correlation functions.}
Following \cite{caputa2017anti}, we search for a profile $\rho(t)$ to minimize the effective action by solving for $\delta W=0$. Eq. \eqref{dW_equal_time} implies that the optimal $\sigma(t)$ obeys the Liouville equation:
\begin{eqnarray} \label{EqualtimeGeom}
\frac{e^{4 \sigma }}{   
    \epsilon  }-\frac{2}{3  
    }\frac{d^2\sigma
   }{\text{dt}^2}=0
 \end{eqnarray}
Much as in \cite{caputa2017anti}, The solution is given by the standard 
substitution of the form $\sigma(t)=-\frac{1}{2}\log  \mu t $, where $\mu=\sqrt{{3} \over{\epsilon}}$ we  find the optimal metric is given by
\begin{eqnarray}
   ds^2=\frac{1}{   
  \mu^2 t^2}\text{dt}^2+\frac{1 }{\mu t }(\text{dx}^2+\text{dy}^2) ~,
\end{eqnarray}
This surprising result suggests that indeed a some type of a hierarchical tensor network would still be the optimal discrete spacetime configuration even if the field theory we started with is only anistropically scale invariant. It is interesting note how the combination $t/\sqrt{\epsilon}$ arises naturally in \eqref{EqualtimeGeom}. This is a natural scaling: If we consider our path integral with action \eqref{ClassicalAction} as describing, e.g. the ground state of the quantum Lifshitz Hamilotnian, and considering the gap scaling as $1\over L^2$ for a system with spatial extent $L$, we see that we would have to evolve the system during time $T\sim L^2$ in order to resolve the low lying states. Setting $\epsilon\sim {1\over L^2}$, we get that the time coordinate has to be scaled as $T\sim {L\over  \sqrt{\epsilon}}$.

Noting that our theory is FPD invariant, it is possible to uniformize the geometry by using a coordinate $u=2\sqrt{t}$ (we take $\epsilon=1$ here),
the optimal metric can also be written as 
\begin{equation}
ds^2 = \frac{4}{3 u^2} \left(du^2 +\sqrt{3} ( dx^2 + dy^2 )\right) ~,
\end{equation}
which is the $AdS_3$ metric of a Poincare patch. Thus, the a proper MERA-like description is possible for this non-uniformally rescaled Lifshitz theory. Another possibility, hinted by recent work on exact holographic tensor networks \cite{alexander2018exact}, is that a non-unitary MERA-like structure may be chosen that features a  scale-invariant tensor network for a non-CFT spin chain model. 

{\bf Optimized geometry for dynamical correlation functions.}.
We turn to address the optimization in the "lateral" direction. In this case our equation is \eqref{dW_dynamical}:
\begin{eqnarray}\label{dynamical_dif_eq1}
 \frac{e^{4 \sigma
   }}{16 \pi \text{  }\epsilon
   }-\frac{e^{2 \sigma
   }(
   \left(\frac{\text{d$\sigma
   $}}{\text{dx}}\right)^2+\frac{d^2\sigma
   }{\text{dx}^2})}{12 \pi
   ^{3/2} \sqrt{\epsilon
   }}=0
\end{eqnarray}
To solve this equation, we define: $Y(x)\equiv e^{\sigma(x)}$, and note that \eqref{dynamical_dif_eq1} can be written as:
\begin{eqnarray}
Y\text{''}=C Y^3~~;~~C=\frac{3\pi ^{1/2}}{4   \sqrt{\epsilon  }
   }
   \end{eqnarray}
This nonlinear equation is equivalent to the system $Y'=Z\text{  };\text{   }Z'={C}{Y^3}$, which allows us to find an integral of motion by solving for $\frac{\text{dZ}}{\text{dY}}=\frac{C Y^3}{Z}$, from which we obtain the integral of motion: 
\begin{eqnarray}
\frac{1}{2} {Y'}^2=C \frac{Y^4}{4}+const.
\end{eqnarray}
We can solve this equation at $const=0$, getting:
\begin{eqnarray}
Y=\frac{\sqrt{2}}{
   \left(\sqrt{C} x+\alpha \right)},
\end{eqnarray}
resulting in the metric, written in terms of $Y$ our metric is 
\begin{eqnarray}
\text{ds}^2=
   Y^4{\text{dt}^2}+  Y^2 ({\text{dx}^2+\text{dy}^2})
\end{eqnarray}
and the leading behavior of the metric at large $x$ is thus:
\begin{eqnarray}\label{LifMet}
\text{ds}^2\approx
   4\frac{\text{dt}^2}{C^2 x^4}+2 \frac{\text{dx}^2+\text{dy}^2}{C x^2}
\end{eqnarray}
We emphasize, that as opposed to the usual notion of holographic Lifshitz geometry for this model, where the boundary is (2+1)-dimensional, here we deform one of the original dimensions of the (2+1) spacetime and use it as our holographic direction. We stress that the geometry \eqref{LifMet} is also suitable for computation of equal point correlation functions, {\it as long as all points involved are along a single line}. On the other hand \eqref{EqualtimeGeom} may be useful for computing any multi point equal correlation functions but not dynamical ones. 

The equal-time and dynamical two-point correlation functions for the quantum Lifshitz model that we consider in this work have been studied in \cite{ardonne2004topological} and more recently in \cite{keranen2017correlation} where they have been compared with the holographic two-point function. The authors find that the correlation functions match quite well with the scaling obtained from a holographic calculation with a Lifshitz geometry, thereby strengthening our expectation that a tensor network description of the system will inherit the features of a Lifshitz geometry.
We find it quite striking that a semi-classical description of correlation functions is obtained for the system, although there is no manifest small parameter like $\hbar$ or a strong/weak coupling duality to drive us into a  semi-classical regime in our original setup.
Finally, we remark that although we obtained here an optimal geometry for a specific $z=2$ (2+1)-dimensional field theory, it is natural to expect that the procedure described here would still work for more general field theories in higher dimensions with arbitrary values of $z$.

%In this regard, we would like to make a few comments. For the case of equal-time two-point correlation function where $\rho = \rho(t)$, our optimized (optimal?) geometry is \textit{different} than the standard vacuum Lifshitz spacetime used in \cite{keranen2017correlation} to evaluate the vaccum two-point function in the geodesic approximation given by
%\begin{equation}
%  \langle \phi(\textbf{r}_1,t_1)   \phi(\textbf{r}_2,t_1)\rangle = |\textbf{r}_1 - \textbf{r}_2 |^{-2\Delta} 
%\end{equation}
%where $\Delta$ is the scaling dimension \footnote{We note that the metric in Eq. 3.32 of \cite{keranen2017correlation} is the same as our optimized geometry in (\ref{EqualtimeGeom}) after their change of variables.}
%\old 

\begin{comment}
\IK 
Todo:: compute in the field theory:
\begin{eqnarray}
\langle \phi(0,0,0)\phi(0,y,t)\rangle
\end{eqnarray}
and the geodesic length $\gamma$ between the points $(x,y,t)=(0,0,0)$ and a different point $(0,y,t)$ in the metric \eqref{LifMet}. When $y,t$ are large, does a relation hold of the form:
\begin{eqnarray}
\langle \phi(0,0,0)\phi(0,y,t)\rangle\propto e^{-a \gamma}
\end{eqnarray}
one can also ask the equivalent question in case (1).
\old
\end{comment}

\emph{Acknowledgments.}
This work was supported in part by the NSF grant DMR-1508245.
\bibliography{LifshitzAnomaly}
\newpage

\onecolumngrid
\appendix

\section{Some details of calculations}
To obtain our equations we carry out a second order perturbation calculation of the heat kernel, using :
\begin{eqnarray}\label{pert_series}
e^{-\epsilon  \left(\rho _0+\delta \rho \right)
   D}=e^{-\tilde{\epsilon } D}-\frac{1}{\rho
   _0}\int _0^{\tilde{\epsilon
   }}e^{-\left(\tilde{\epsilon }-s\right)\text{ 
   }D} \delta \rho  D\text{  }e^{-s\text{ 
   }D}\text{ds}+\frac{1}{\rho _0^2}\int
   _0^{\tilde{\epsilon }}\text{ds} \int
   _0^s\text{ds}_1e^{-\left(\tilde{\epsilon
   }-s\right)\text{  }D} \delta \rho  D\text{ 
   }e^{-\left(s-s_1\right)\text{  }D} \delta \rho
    D\text{  }e^{-s_1 D}
\end{eqnarray}
where $\tilde{\epsilon }=\rho _0\epsilon$.
For convenience, set ${\bf r}_0=0$ throughout the calculation, and reinstate its value in the end. We assume that the operator $D$ is diagonal in momentum, and that $\delta\rho$ depends on a single coordinate $x$, and has an expansion:
\begin{eqnarray}
\delta \rho =\Sigma _{m=1} c_m x^m
\end{eqnarray}
Taking $q$ to be the momentum in the $x$ direction and $K$ to be the momentum vector in all other directions, the zeroth order contribution to the heat kenrel reads:
\begin{eqnarray}
A_0=\langle 0|e^{-\tilde{\epsilon } D}|0\rangle
   =\frac{1}{(2\pi )^{d+1}}\int d^dK
   \text{dq}\text{  }e^{-\tilde{\epsilon }
   D(K,q)}\text{   };\text{   }
\end{eqnarray}
The contribution from the first order term in \eqref{pert_series} is
\begin{eqnarray}&
A_1=-\frac{1}{\rho
   _0}\langle 0|\int _0^{\tilde{\epsilon
   }}e^{-\left(\tilde{\epsilon }-s\right)\text{ 
   }D} \delta \rho  D\text{  }e^{-s\text{ 
   }D}|0\rangle\text{ds}=\\ \nonumber &-\frac{1}{\rho _0}\frac{2\pi }{(2\pi
   )^{d+2}}\int _0^{\tilde{\epsilon }}
   \text{ds}\int d^dK \text{dq} \left(\Sigma 
   c_m\left(i\frac{d}{d
   q}\right){}^me^{-\left(\tilde{\epsilon
   }-s\right)\text{  }D(K,q)}\right) D(K,q)\text{
    }e^{-s\text{   }D(K,q) }
\end{eqnarray}
which can also be expressed in the form:
\begin{eqnarray}&
A_1=\\ \nonumber & -\frac{1}{\rho _0}\frac{1}{(2\pi
   )^{d+1}}\int _0^{\tilde{\epsilon }}
   \text{ds}\int d^dK \text{dq} e^{-\tilde{\epsilon } D(K,q)
   } D(K,q)\{\Sigma
   _{m=1} i^m\Sigma
   _{h=1}^m(-1)^h c_m B_{h,m}\left(\left(\tilde{\epsilon
   }-s\right)D'\left(K,q_1\right),(\tilde{\epsilon
   }-s)D\text{''}\left(K,q_1\right),\text{
   ...}\right) \}
\end{eqnarray}
where $B_{h,m}$ are Bell polynomials. In the case we are interested in, due to the time reversal/space inversion symmetry the first non zero contribution comes from $c_2={1\over 2}{\partial_x}^2 \delta\rho$:
\begin{eqnarray}
A_1\approx \frac{1}{\rho _0}\frac{c_2 }{(2\pi
   )^{d+1}}\int d^dK \text{dq}
   e^{-\tilde{\epsilon }\text{   }D(K,q)
   }\left(-\frac{1}{2} D\text{''}(K,q)
   \tilde{\epsilon }^2+\frac{1}{3} (D'(K,q))^2
   \tilde{\epsilon }^3\right)
\end{eqnarray}
The second order contribution is given by:
\begin{eqnarray}&
A_2=\frac{1}{\rho _0^2}\langle 0|\int
   _0^{\tilde{\epsilon }}\text{ds} \int
   _0^s\text{ds}_1e^{-\left(\tilde{\epsilon
   }-s\right)\text{  }D} \delta \rho  D\text{ 
   }e^{-\left(s-s_1\right)\text{  }D} \delta
   \rho  D\text{  }e^{-s_1 D}|0\rangle
   =\\ \nonumber & \frac{1}{\rho _0^2}\frac{\Sigma
   _{n,m}c_n c_m}{(2\pi )^{d+1}}\int
   _0^{\tilde{\epsilon }}\text{ds} \int
   d^dK\text{  }\text{dq}
   \left(\left(i\frac{d}{d
   q}\right)^me^{-\left(\tilde{\epsilon
   }-s\right)\text{  }D(K,q)}\right)D(K,q)\text{
    }e^{-\left(s-s_1\right)\text{   }D(K,q)
   }\left(\left(-i\frac{d}{d
   q}\right)^nD(K,q)\text{  }e^{-s_1\text{ 
   }D(K,q) }\right)
\end{eqnarray}

\end{document}